\documentclass[journal=jacsat,manuscript=article]{achemso}

\usepackage[version=3]{mhchem} 

\author{Chong Ye}
\affiliation{Key Laboratory of advanced optoelectronic quantum architecture and measurements of Ministry of Education, Beijing Key Laboratory of Nanophotonics $\&$ Ultrafine Optoelectronic Systems, School of Physics, Beijing Institute of Technology, 100081, Beijing, China}
\author{Yifan Sun}
\affiliation{Key Laboratory of advanced optoelectronic quantum architecture and measurements of Ministry of Education, Beijing Key Laboratory of Nanophotonics $\&$ Ultrafine Optoelectronic Systems, School of Physics, Beijing Institute of Technology, 100081, Beijing, China}
\author{Xiangdong Zhang}
\affiliation{Key Laboratory of advanced optoelectronic quantum architecture and measurements of Ministry of Education, Beijing Key Laboratory of Nanophotonics $\&$ Ultrafine Optoelectronic Systems, School of Physics, Beijing Institute of Technology, 100081, Beijing, China}
\email{zhangxd@bit.edu.cn}

\title[An \textsf{achemso} demo]
  {Entanglement-Assisted Quantum Chiral Spectroscopy}

\abbreviations{IR,NMR,UV}

\begin{document}

\begin{tocentry}

Some journals require a graphical entry for the Table of Contents.
This should be laid out ``print ready'' so that the sizing of the
text is correct.

Inside the \texttt{tocentry} environment, the font used is Helvetica
8\,pt, as required by \emph{Journal of the American Chemical
Society}.

The surrounding frame is 9\,cm by 3.5\,cm, which is the maximum
permitted for  \emph{Journal of the American Chemical Society}
graphical table of content entries. The box will not resize if the
content is too big: instead it will overflow the edge of the box.

This box and the associated title will always be printed on a
separate page at the end of the document.

\end{tocentry}

\begin{abstract}
  The most important problem of spectroscopic chiral analysis is the inherently weak chiral signals are easily overwhelmed by the environment noises. Enormous efforts had been spent to overcome this problem by enhancing the symmetry break in the light-molecule interactions or reducing the environment noises. Here, we propose an alternative way to solve this problem by using frequency-entangled photons as probe signals and detecting them in coincidence, i.e., using quantum chiral spectroscopy. For this purpose, we develop the theory of entanglement-assisted quantum chiral spectroscopy. Our results show that the signals of left- and right-handed molecules in the quantum spectrum are always distinguishable by suitably configuring the entangled probe photons. In construct, the classical spectrum of the two enantiomers become indistinguishable when the symmetry break in the interactions is overwhelmed by the environment noises. This offers our quantum chiral spectroscopy a great advantage over all classical chiral spectroscopy. Our work opens up an exciting area that exploring profound advantages of quantum spectroscopy in chiral analysis.
\end{abstract}

\section{Introduction}
Chiral molecules cannot be superimposed with their mirror images by rotations and translations.
Such stereo-isomers are called enantiomers. As with our hands, the two enantiomers can be termed by
left-handed and right-handed molecules, respectively. The enantiomers of opposite chirality (handedness)
play dramatically different roles in chemistry, biotechnologies, and pharmaceutics~\cite{A1}. Because the two enantiomers share almost all of their physical properties, identifying them is challenging and of great importance to the above and other relevant research areas~\cite{QK,HL}.

In the passing years, spectroscopy methods of chiral analysis are developed to distinguish molecular chirality. The main idea of the methods is to induce the symmetry break in the interactions between light and two enantiomers so that the chiral difference can be mapped to the optical degrees of freedom and detected spectroscopically. By far, well established spectroscopy methods of chiral analysis include (vibrational~\cite{B2} and photoelectron~\cite{beaulieu2018photoexcitation}) circular dichroism~\cite{B1}, optical rotary dispersion~\cite{B0}, Raman optical activity~\cite{B7}, and recent three-wave mixing spectroscopy~\cite{De1}.

However, the molecular size is typically smaller
than the wavelengths of the applied electromagnetic fields, leading to inherently weak chiral signals. Such weak chiral signals are easily overwhelmed by the environment noises, which affect the chiral analysis in the process of light-molecule interactions and the detection process. In the first process, the symmetry break in the interactions is suppressed by the environment noises, making it difficult to map the chiral difference to optical degrees of freedom. In the second process, the environment noises make it hard to detect the chiral signals. In order to solve this problem, enormous efforts had been made to reduce the environment noises by using buffer-gas cooling~\cite{CR,MP}, and direct laser cooling~\cite{SCD,PRXD}. Alternatively, advancements toward enhanced the symmetry break in light-molecule interactions have been widely explored, such as using plasmonic nanostructures~\cite{Zhang3}, chiral surface~\cite{PRL.118.193401}, and twisted light with optical orbital angular momentum~\cite{F3}.

In this Letter, we propose a quantum chiral spectroscopy to solve such a problem by using a more ingenious
detecting scheme. To this end, we develop the theory of using frequency-entangled photons as probe signals and detecting them in coincidence. Attributing to the nonclassical bandwidth of
the frequency-entangled photons~\cite{M1,M2,M3}, the chiral difference can be enhanced in the coincidence signals. Our results show that the signals of left- and right-handed molecules in the coincidence spectrum are always distinguishable by suitably choosing the frequency-entangled two-photon pairs. In contrast, the classical spectrum left- and right-handed molecules will become indistinguishable in the strong dissipation region, where the symmetry break in the interactions is overwhelmed by the environment noises. This offers our quantum chiral spectroscopy a great advantage over all classical chiral spectroscopes.

\section{Model}
\begin{figure}[ht]
	\centering
	\includegraphics[width=0.8\columnwidth]{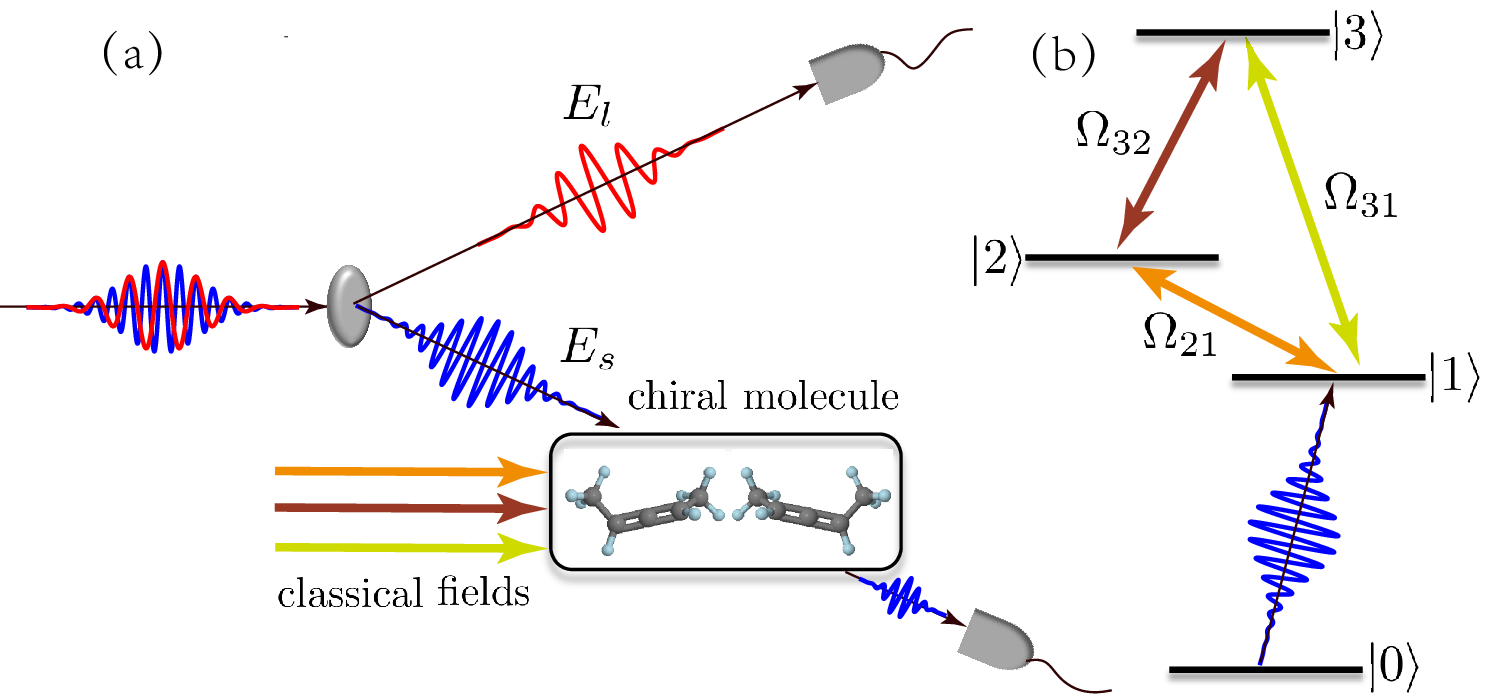}\\
	\caption{(a) Proposed setup for chiral analysis: Three classical fields are offered to control the chiral molecule. $E_s$ and $E_l$ are frequency-entangled. While $E_s$ interacts with the chiral molecule, $E_l$ propagates freely. The two photons are detected in coincidence. (b) The level scheme of our working system without regard to its chirality. The molecular chirality is reflected in the coupling strengths $\Omega_{ij}$.}\label{Fig0}
\end{figure}
Our theoretical model is shown in Fig.~\ref{Fig0}\,(a), where a chiral molecule couples with three narrow-band classical control fields and the signal photon of an incident broad-band entangled two-photon pair through the electric dipole interactions. A detector is used to collect the transmitted signal photon. The idle photon will not interact with the chiral molecule and serves as a reference for the two-photon counting. We use a four-level model with a cyclic three-level configuration as shown in Fig.~\ref{Fig0}\,(b) to describe our working system. The three excited states $|1\rangle$, $|2\rangle$, and $|3\rangle$ are coupled to each other in a cyclic three-level configuration through the three classical control fields. The ground state $|0\rangle$ is coupled to the excited state $|1\rangle$ through the signal photon.

The energies of molecular states $|j\rangle$ are $\hbar\omega_{j}$. The frequencies of the three narrow-band control fields are $v_{ij}$ ($i>j=1,2,3$).
The corresponding coupling strengths are $\Omega_{ij}$.
For simplicity, the system is set to work under the
the three-photon resonance condition $v_{31}=v_{21}+v_{32}$.
For the broad-band signal and idle photons in our setup, we assume the paraxial and long wave
approximations such that they can be simplified to be one-dimensional~\cite{M1,M3}, respectively.

In the rotating-wave approximation, the Hamiltonian of the four-level model at $r=0$ without regard to the molecular chirality is ($\hbar=1$)
\begin{align}\label{HM}
{H}&=\sum^{3}_{j=0}\omega_{j}\sigma_{jj}
+\sum^{3}_{i>j=1}(\Omega_{ij}\sigma_{ij}e^{-\mathrm{i}v_{ij}t}+\mathrm{H.c.})+
\sum_{k_s}[\omega_{s}a^{\dagger}_{k_s}a_{k_s}+(g_{k_s}a_{k_s}\sigma_{10}+
\mathrm{H.c.})]+\sum_{k_l}\omega_{l}a^{\dagger}_{k_l}a_{k_l}.
\end{align}
Here, we have defined $\sigma_{ij}\equiv|i\rangle\langle j|$ with $i,j=0,1,2,3$.
$a^{\dagger}_{k_s}$ and $a^{\dagger}_{k_l}$ are the creation operators for the signal and
idle photons with energies $\hbar\omega_s$ and $\hbar\omega_l$. The coupling strength attributing to the signal photon is $g_{k_s}=\mu\sqrt{\omega_{s}/(2\varepsilon_0L)}$.
Here, $\mu$, $L$, and $\varepsilon_0$ are the transition electric dipole, the quantized volume, and the dielectric constant in the vacuum, respectively.

The chirality of our working system can survive in the electric dipole approximation. It is reflected by the three electric dipole transition moments of the cyclic three-level configuration. While their magnitudes are the same for the two enantiomers, the sign of their product is distinct for each enantiomer~\cite{T1,ye2018real}. This property plays the essential role in chiral analysis~\cite{De1,ye2019determination,Pr10} and
a more ambitious issue named enantio-purification~\cite{Pr1,Pr5,ye2019effective,ye2020fast,ye2021enantio,Pe1,Pe2}. In our four-level model, this property is revealed by the sign of the coupling strengths $\Omega_{ij}$. Without loss of generality, we can choose
\begin{align}\label{CDE}
\Omega^{R}_{31}=-\Omega^{L}_{31},~~\Omega^{R}_{21}=\Omega^{L}_{21},~~\Omega^{R}_{32}=\Omega^{L}_{32}.
\end{align}
Here, the subscripts ``$Q=L,R$" are used to denote the molecular chirality.

\begin{figure}[ht]
	\centering
	\includegraphics[width=0.9\columnwidth]{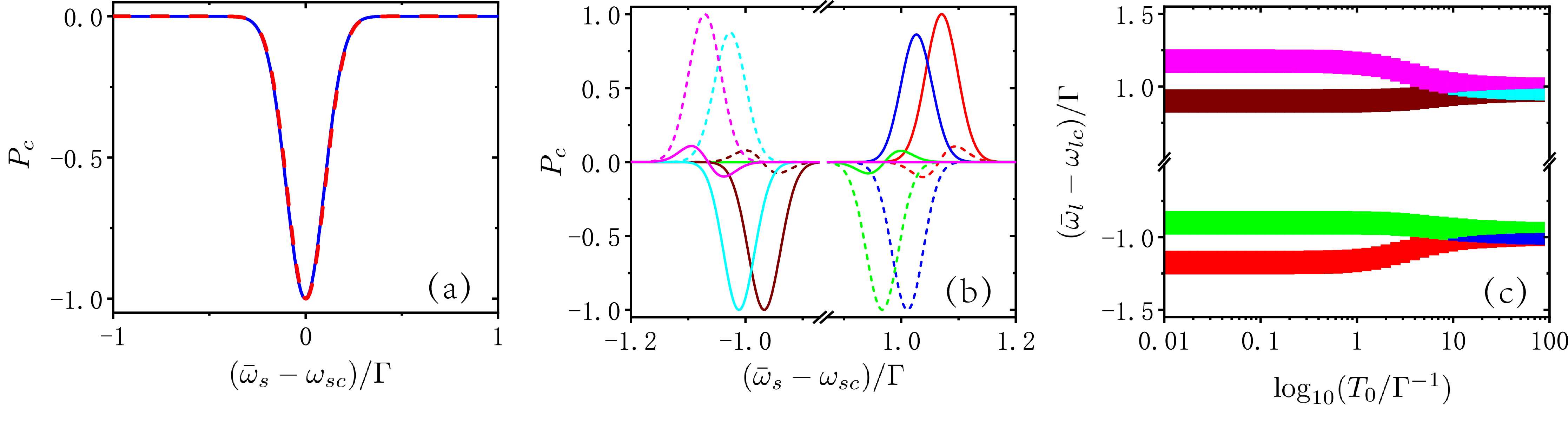}\\
	\caption{Classical (a) vs Quantum (b) chiral spectroscopy of left-handed (solid line) and right-handed (dashed line) molecules in strong dissipation region. In quantum chiral spectroscopy, we give six cases (distinguished by colors) corresponding to different $\bar{\omega}_{l}$. The spectrum of the two enantiomers have different spectral-line shapes or signs. (c) The working regimes (colored regimes) of our entanglement-assisted quantum chiral spectroscopy in the $(T_0-\bar{\omega}_l)$ plane. Each regime in (c) corresponds to one of the six cases in (b).  }\label{Fig2}
\end{figure}

\section{Frequency-resolved chiral coincidence spectrum}
In our setup, the chiral signals are collected by two-photon counting and given by
coincidence probabilities in the standard Glauber's approach~\cite{Glauber}.
We are interested in the coincidence probabilities of detecting
a single photon at $r_s$ and a single idle photon at $r_l$ is detected.
We assume that the sensitive function of the two detectors are $\mathcal{S}\delta(\bar{\omega}_{s}-\omega_{s})$ and $\mathcal{S}\delta(\bar{\omega}_{l}-\omega_{l})$ with
sensitive frequencies $\bar{\omega}_{s}$ and $\bar{\omega}_{l}$. Then, by fixing the sensitive frequency of one detector and
scanning that of the other, we can obtain the frequency-resolved spectrum of the coincidence probability, i.e., the frequency-resolved coincidence spectrum.

In the standard Glauber's approach~\cite{Glauber}, the coincidence probabilities can be achieved with the help of the system's evolution operators, which can be obtained by using the following procedure (see more details in Section 1 of Supplement). We treat the photon-molecule coupling $\sum_{k_s}(g_{k_s}a_{k_s}\sigma_{10}+\mathrm{H.c.})$ as a perturbation. The remaining terms can be diagonalized in the interaction picture as $\mathcal{H}^{(0)}=\sum^{3}_{i=1}\lambda^{Q}_{i}|\lambda^{Q}_{i}\rangle\langle \lambda^{Q}_i|+
\sum_{k_s}\Delta_{k_s}a^{\dagger}_{k_s}a_{k_s}+\sum_{k_l}\omega_{l}a^{\dagger}_{k_l}a_{k_l}$ with $\Delta_{k_s}\equiv \omega_{s}-(\omega_{1}-\omega_0)$. Using Dyson series, we
give the evolution operators of the system and the coincidence spectrum. The coincidence spectrum are composed of two parts. One is the background, which describes
the frequency-resolved coincidence probabilities without the chiral molecule. The other is the frequency-resolved
transmission spectrum~\cite{M2}, which is the change of the frequency-resolved coincidence probabilities due to the appearance of the chiral molecule.

Since $\lambda^{Q}_i$ are chirality-dependent, the transmission spectrum are different for the two enantiomers.
Specifically, for the system initially prepared in
\begin{align}\label{HES}
|\Psi(0)\rangle=|0\rangle\otimes
\sum_{k_s,k_l}\psi(k_s,k_l)a^{\dagger}_{k_s}
a^{\dagger}_{k_l}|vac\rangle
\end{align}
with the vacuum state of the photon field $|vac\rangle$, the transmission spectrum are (see more details in Section 2 of Supplement)
\begin{align}\label{TSC}
P^{Q}_{c}(\bar{\omega}_{s},\bar{\omega}_{l})
=&-\frac{\mathcal{S}^2L^3}{8\pi^4}\varepsilon^2_{\bar{k}_l}\varepsilon^2_{\bar{k}_s}
\sum^{3}_{i=1}\sum_{k^{\prime\prime}_s}|\eta_{1\lambda^{Q}_i}|^2
\Re\left[\frac{g^{\ast}_{\bar{k}_s}g_{k^{\prime\prime}_s}
\psi^{\ast}(\bar{k}_s,\bar{k}_l)\psi(k^{\prime\prime}_s,\bar{k}_l)}
{(\lambda^{Q}_{i}-\Delta_{\bar{k}_s}+\mathrm{i}\Gamma)(\lambda^{Q}_{i}-\Delta_{k^{\prime\prime}_s}+\mathrm{i}\Gamma)}
\right]
\end{align}
with $\varepsilon_{k_s}=\sqrt{\omega_{s}/(2\varepsilon_0L)}$, $\varepsilon_{k_l}=\sqrt{\omega_{l}/(2\varepsilon_0L)}$,
$\bar{k}_s=\bar{\omega}_s/c$, $\bar{k}_l=\bar{\omega}_l/c$, and $\eta_{1\lambda^{Q}_i}=\langle 1|\lambda^{Q}_i\rangle$.  $c$ is the speed of
light in vacuum.
The environment noises will affect the process of the light-molecule interactions, which give rise to the dissipation of molecules. In order to describe the effect of dissipation, we have introduced $\Gamma$ in
Eq.~(\ref{TSC}).

\section{Entanglement-enhanced chiral analysis}
In general, the chirality dependency in the transmission spectrum is independent of whether the two photons are entangled or not. However, in the classical chiral spectroscopy, where the two uncorrelated photons are used, the spectral broadening caused by dissipation will reduce the chirality dependency and eventually make the transmission spectrum of the two enantiomers become indistinguishable. In contrast, in quantum chiral spectroscopy, where frequency-entangled photons are used as probe signals, the signal can be filtered and thus the chirality dependency in the transmission spectrum can be enhanced.

In order to demonstrate such an advantage of quantum chiral spectroscopy, we set the chiral molecule in the strong dissipation region
by choosing the coupling strengths for the two enantiomers as
$\Omega^{R}_{31}=-\Omega^{L}_{31}=0.1\Gamma$, $\Omega^{R}_{21}=\Omega^{L}_{21}=0.1\Gamma$, and
$\Omega^{R}_{32}=\Omega^{L}_{32}=0.1\Gamma$.
The three classical control fields are coupled with their corresponding transitions on resonance with $\Delta_{21}=\Delta_{31}=0$.
Under these parameters, the classical transmission spectrum of the two enantiomers are
indistinguishable as shown in Fig.~\ref{Fig2}(a). For calculations, we have used uncorrelated two-photon pairs with center frequencies $\omega_{sc}$ and $\omega_{lc}$ by choosing $\psi(k_s,k_l)\propto\exp\{-[(\omega_{s}-\omega_{sc})^2+(\omega_{l}-\omega_{lc})^2]/(2\sigma^2)\}$ with width $\sigma=\Gamma$.

For the case of frequency entangled two-photon pairs, we use~\cite{M2}
\begin{align}\label{EQ5}
\psi(k_s,k_l)\propto A_{p}(k_{s},k_{l})\exp{(-\kappa^2)}.
\end{align}
Such two-photon pairs can be generated in nonlinear crystals due to the downconversion process~\cite{M3}.
The envelope $A_{p}(k_{s},k_{l})$ is given by the pumped pulse. Here, we use a Gaussian envelop with $A_{p}(k_{s},k_{l})=\exp{[-{(\omega_{l}+\omega_{s}-\omega_p)^2}/{(2\sigma_p^2)}]}$
with the center frequency $\omega_p$ and the width $\sigma_p$.
The phase-mismatching function $\exp{(-\kappa^2)}$ appearances due to the propagation of photons in the crystals. For entangled two-photon pairs generated in type-II nonlinear crystals, we approximately have~\cite{M1,M2,M3}
\begin{align}\label{EQ6}
\kappa=(\omega_s-\omega_{sc})\frac{T_s}{2}+
(\omega_l-\omega_{lc})\frac{T_l}{2}
\end{align}
with the center frequencies of the two photons $\omega_{sc}$ and $\omega_{lc}$. $T_s$ and $T_l$ are the maximal time delays of
the two photons propagating in the crystals with respect to the pumped pulse.

In Fig.~\ref{Fig2}(b), we show the transmission spectrum of the two enantiomers in the strong dissipation region by using entangled two-photon pair with $\sigma_p=\Gamma$, $T_s=2.4\times 10^1\Gamma^{-1}$, and $T_l=2.5\times10^1\Gamma^{-1}$. By tuning  $\bar{\omega}_l$, the transmission spectrum of the two enantiomers with respect to $\bar{\omega}_s$ can have different spectral-line shapes or signs. Here, we have illustrated six cases distinguished by color. Comparing these results with that in Fig.~\ref{Fig2}(a), we can conclude that the chiral analysis using frequency entangled two-photon pairs is overwhelming advantage over that using uncorrelated ones in the strong dissipation region. The advantage is due to the appearance of frequency entanglement. In this sense, we have demonstrated an entanglement-assisted quantum chiral spectroscopy in the strong dissipation region.

In addition to $\bar{\omega}_l$, $T_s$ and $T_l$ are other two additional tunable parameters in quantum spectroscopy comparing with the classical spectroscopy. It is natural
to explore the working regimes of our entanglement-assisted quantum chiral spectroscopy with respect to $\bar{\omega}_l$, $T_s$ and $T_l$. For simplicity, we have used $T_s=2.4T_0$ and $T_l=2.5 T_0$. The results are shown in Fig~\ref{Fig2}(c). Our entanglement-assisted quantum chiral spectroscopy works in the colored regimes of the $(T_0-\bar{\omega}_l)$ plane. Each regime in Fig.~\ref{Fig2}(c) corresponds to one of the six cases in Fig.~\ref{Fig2}(b) and is marked by the same color. These results indicate that the entanglement-assisted quantum chiral spectroscopy is robust against the variations of the entangled two-photon pairs and the detector.

\section{Physical role of correlation}
The frequency correlation between the two photons plays an important role in our entanglement-assisted quantum chiral spectroscopy.
In order to illustrate this explicitly, we consider a photon pair with a zero-bandwidth negative energy correlation, given by~\cite{M3}
\begin{align}
	\psi(k_s,k_l)=\phi_s(k_s)\delta(\omega_{l}+\omega_{s}-\omega_p).
\end{align}
The delta function $\delta(\omega_{l}+\omega_{s}-\omega_p)$ indicates the
the energy conservation of two photons.
Moreover, we choose
$\Delta_{12}=\Delta_{13}\gg \{|\Omega_{12}|,|\Omega_{13}|\}$. In this case, to the second order, the dressed state $|\lambda^{Q}_{1}\rangle\simeq |1\rangle$~\cite{Pr1}, i.e., $\eta_{1\lambda^{Q}_2}\simeq 0$ and $\eta_{1\lambda^{Q}_3}\simeq 0$.
The transmission spectrum can be greatly simplified as
\begin{align}\label{DEP}
	P^{Q}_{c}(\bar{\omega}_{s},\bar{\omega}_{l})
	=&-\frac{\mathcal{S}^2L^4}{16\pi^4}\varepsilon^2_{\bar{k}_l}\varepsilon^2_{\bar{k}_s}
	\delta(\bar{\omega}_{l}+\bar{\omega}_{s}-\omega_p)|\eta_{1\lambda^{Q}_i}|^2\Re\left[\frac{g_{\bar{k}_s}g_{\bar{k}_{pl}}\phi^{\ast}_s(\bar{k}_s)\phi_s(\bar{k}_{pl})}
	{(\lambda^{Q}_{1}-\Delta_{\bar{k}_s}+\mathrm{i}\Gamma)
(\lambda^{Q}_{1}-\Delta_{\bar{k}_{pl}}+\mathrm{i}\Gamma)}
	\right]
\end{align}
with $\bar{k}_{pl}\equiv k_p-\bar{k}_l$. Due to the delta function in $\psi(k^{\prime\prime}_s,\bar{k}_l)$, only the term of $k^{\prime\prime}_s=k_p-\bar{k}_l$ with $k_p=\omega_p/c$ is nonzero and will contribute to the summation over $k^{\prime\prime}_s$ in Eq.~(\ref{TSC}), i.e., the correlation filters our signals.

The entanglement-assisted quantum chiral spectroscopy is also robust against the variation of dissipation.
For a fixed $\bar{\omega}_l$, $P_c$ is nonzero at $\bar{\omega}_s=\omega_p-\bar{\omega}_l$, i.e.,
$\bar{k}_{s}=\bar{k}_{pl}$. The sign of $P^{Q}_c$ is determined by the sign of $[(\lambda^{Q}_{1}-\Delta_{\bar{k}_{pl}})^2-\Gamma^2]$. Since $\lambda^{L}_{1}\ne \lambda^{R}_1$~\cite{Dt9},
$P^{Q}_c$ have different signs for the two enantiomers when $\Delta_{\bar{k}_{pl}}=
\omega_p-\bar{\omega}_l-(\omega_1-\omega_0)$ is in the range between
$(\Gamma-\lambda^{L}_{1})$ and $(\Gamma-\lambda^{R}_{1})$. Since the photon
pair with a zero-bandwidth negative energy correlation can be approximately achieved by adjusting the
width of the pumped pulse and the delay times~\cite{M3}, the transmission spectrum of the two enantiomers are always
distinguishable in the quantum chiral spectroscopy by suitable choosing the entangled photons.

\section{Discussions and Conclusions}
In summary, we have developed the theory of entanglement-assisted quantum chiral spectroscopy to detect the molecular chirality. It offers an alternative way for suppressing the effect of environment noises on
chiral analysis. Using a theoretical model, we have demonstrated that
attributing to the nonclassical bandwidth of
the frequency-entangled photons, the chiral difference can be enhanced in the coincidence signals.
It can be applied
to distinguish the molecular chirality in the strong dissipation region, where the classical chiral spectroscopy based on uncorrelated light loses efficiency due to the dissipation-induced spectral broadening.

As an experimentally accessible approach, frequency-entangled photons can be produced by spontaneous parametric down conversion. The apparatuses for generating the required entangled photons in our discussions are actually available. For example, when the transition $|0\rangle\leftrightarrow|1\rangle$ is between vibrational levels, the signal photon is in the optical domain, which has been broadly investigated~\cite{Prabhakar2020}. Also, the cyclic three-level configuration can be realized in the optical domain~\cite{Pr5}, which means the whole procedure is workable in such a frequency domain.

Quantum spectroscopy that using frequency-entangled photons as probe signals is gaining attentions due to unique features of frequency-entangled photons, such as quantum fluctuations~\cite{QL1}, linear scaling behaviors~\cite{QL3}, and nonclassical bandwidth~\cite{M1,M2,M3}.
We for the first time introduce it to chiral analysis and
demonstrate the advantage of using frequency-entangled photons as probe signals due to their
nonclassical bandwidth.
Despite this, quantum spectroscopy had been
demonstrated to achieve technical benefits, such as probing of the enormous range of wavelengths using visible photons~\cite{RM1,RM2}, to offer the possibility of cancelling uncorrelated background optical noises, and
noises of the detectors~\cite{TS2}, and to achieve a provable quantum advantage over all schemes using classical sources~\cite{M4}. Our work opens up an exciting area that exploring the profound advantages of quantum spectroscopy in chiral analysis.

\begin{acknowledgement}
The authors thank the support by
National Natural Science Foundation of China (No. 91850205), National Science Foundation for Young Scientists of China (No.11904022), and Beijing Institute of Technology Research Fund Program for Young Scholars.

\end{acknowledgement}

\begin{suppinfo}
Derivation of frequency-resolved chiral coincidence signals.
\end{suppinfo}

\bibliography{achemso-demo}

\end{document}